\documentclass[aps,prl,twocolumn]{revtex4-1}	
\usepackage{graphicx}
\usepackage{amsmath, amsfonts,amssymb}
\usepackage{color}

\newcommand{\be}{\begin{equation}}
\newcommand{\ee}{\end{equation}}

\newcommand{\Tr}{{\rm Tr}}

\begin{document}
\title{Response to ``Comment on {Universal Lindblad Equation for open quantum systems}''}
\author{\vspace{-0.05 in}Frederik Nathan and Mark S. Rudner}
\affiliation{Niels Bohr Institute, University of Copenhagen, 2100 Copenhagen, Denmark}
\begin{abstract}
  \vspace{-0.05 in} In a recent comment, 
Lee and Yeo  
   show that the Gibbs state is not generically an exact steady state of the Universal Lindblad Equation (ULE) that we developed in Phys.~Rev.~B {\bf 102}, 115109 (2020). 
  This  non-controversial  observation is 
   precisely as expected for open quantum systems with finite system-bath coupling, where transition rates may be comparable to or
    larger than the level spacing of the system, and we made no claim to the contrary in our paper.
The 
   comment by Lee and Yeo
 hence  highlights 
  that the ULE 
   captures  {
 contributions to the 
  steady state 
 due to finite system-bath coupling} 
     that are beyond the reach of master equations that rely on rotating wave approximations.
 In this response we further clarify the nature of our 
   analytical and numerical results. 
\end{abstract}
\maketitle

In Ref.~\cite{Nathan_2020} we derive a ``universal Lindblad equation'' (ULE), which is
a Markovian master equation that can be used to describe the evolution of a wide variety of open quantum few and many-body systems. 
The regime of validity of the ULE is defined solely in terms of an intrinsic correlation time of the {bath}, and the {system-bath coupling}, without requiring any restrictions on the internal energy level structure of the system itself.
 In particular, the ULE does not rely on the rotating wave (i.e., secular) approximation, and therefore can be applied to systems where the level spacing is small compared with relaxation rates that arise due to the system-bath coupling~\cite{Nathan_2020}.

In their recent comment~\cite{Lee_2020}, Lee and Yeo set out to demonstrate that when a system with Hamiltonian $H_{\rm S}$ is connected to a bath in thermal equilibrium at inverse temperature $\beta$, the Gibbs state $\rho_{\rm G}\equiv e^{-\beta H_{\rm S}}/\Tr(e^{-\beta H_{\rm S}})$ is not an exact steady state of the ULE.
Comment~\cite{Lee_2020}  appears to present this observation as either a deficiency of the ULE or as a contradiction to the claims of our paper; however, it is neither.
On basic physical grounds, it is expected that the steady state of an open system should deviate from $\rho_{\rm G}$ when the system-bath coupling is finite (see, e.g., Ref.~\cite{Thingna_2012} and references therein).
For this reason we did not claim that the steady state should generically be given by $\rho_{\rm G}$, as suggested in Ref.~\cite{Lee_2020}.
To avoid potential future misunderstanding on this point, here we further clarify the nature of our results. 

{\it Steady states of open quantum systems ---}
As noted in Comment~\cite{Lee_2020}, it is straightforward to verify that the Gibbs thermal state $\rho_{\rm G}$ is not generically a steady state of the ULE by direct insertion into Eqs.~(1)-(2) of Ref.~\cite{Nathan_2020}.
The one case where the steady state of the ULE does become identical to $\rho_{\rm G}$ 
is the limit where the effective system-bath coupling $\Gamma$ is much smaller than the level spacing of the system.
As we explain in Sec.~I of Ref.~\cite{Nathan_2020}, in this limit the ULE reduces to the quantum optical master equation (or conventional Lindblad equation as referred to in Ref.~\cite{Lee_2020}).
Evidently, in the limit $\Gamma \to 0$ the steady state of the ULE is thus identical to $\rho_{\rm G}$ (as was also found in Ref.~\cite{Lee_2020}).
The key advance of the ULE is that it applies both within and outside the regime where the rotating wave approximation (and hence the quantum optical master equation) is valid. 
We hence expect the ULE  to faithfully describe finite-coupling corrections to the steady state when $\Gamma$ is large compared to the system’s level spacing, as long as the product of $\Gamma$ and the correlation time of the bath remains small (i.e., $\Gamma \tau \ll 1$ in the notation of Ref.~\cite{Nathan_2020}).

{\it Role of numerical simulations ---}
The role of the numerical simulations in Sec.~V of Ref.~\cite{Nathan_2020} was to illustrate 
 how the ULE could be applied to a quantum many-body system (a chain of 12 spins) where, due to the small level spacing of the system and moderate system-bath coupling, the 
quantum optical master equation could not be used. 
There, we first studied a situation where the bath was only connected to a single site at the end of a chain.
In this case it is naturally expected that, in the steady state, local observables with support far from the point of contact with the bath will take values that are indistinguishable from those in thermal equilibrium. 
We thus computed the bulk magnetization density $\langle M\rangle$  as a ``sanity check'' and point of reference to demonstrate that the steady state of the ULE for an equilibrium setting is in this sense locally indistinguishable from the Gibbs state. 
This was what we meant by the statement ``the universal Lindblad equation reproduces the expected equilibrium steady states.''
The statement was not intended to imply that the steady state (many-body) density matrix of the ULE should be precisely $\rho_{\rm G}$. 

We thank the authors of Ref.~\cite{Lee_2020} for identifying three typos in Appendix D of Ref.~\cite{Nathan_2020}, which we will correct.
As  discussed in the final remarks of Ref.~\cite{Lee_2020}, it will be an interesting direction for future work to investigate the steady states of the ULE, including deviations from Gibbs states arising due to finite system-bath coupling.

 
%
%
\vspace{-0.2 in}
\bibliographystyle{apsrev}
\bibliography{Comment_bibliography.bib}
\end{document}